\documentclass[a4paper,12pt]{article}
\usepackage{jheppub,esint,shuffle,psfrag}
 \usepackage[utf8]{inputenc}

\usepackage{esint} 
\usepackage{breqn}

\def \tr {\mathop{\rm tr}\nolimits}

\def \e  {\mathop{\rm e}\nolimits}

\newcommand\lr[1]{{\left({#1}\right)}}

\newcommand \vev [1] {\langle{#1}\rangle}

\newcommand \ket [1] {|{#1}\rangle}

\newcommand\re[1]{(\ref{#1})}

\def \qqqquad {\qquad\qquad}

\newcommand{\ft}[2]{{\textstyle\frac{#1}{#2}}}

\def\var{\epsilon}

\def\numberbysection{\@addtoreset{equation}{section}
                     \def\theequation{\thesection.\arabic{equation}}}

\title{On level crossing in conformal field theories}
\author {G.P.~Korchemsky}
\affiliation {Institut de Physique Th\'eorique\footnote{Unit\'e Mixte de Recherche 3681 du CNRS}, CEA Saclay, 91191 Gif-sur-Yvette Cedex, France}
\preprint{  \parbox[t]{28mm}{IPhT-T15/213}}

 \abstract{We study the properties of operators in a unitary conformal field theory whose scaling dimensions approach each other for some
 values of the parameters and satisfy von Neumann--Wigner non-crossing rule. We argue that the scaling dimensions of such operators and 
 their OPE coefficients have a universal scaling behavior in the vicinity of the crossing point. We demonstrate that the obtained relations are in 
 a good agreement with the known examples of the level-crossing phenomenon in maximally supersymmetric $\mathcal N=4$ Yang-Mills
 theory,   three-dimensional conformal field theories and QCD.}

\begin{document}
 
\maketitle

\flushbottom
 
\section{Introduction}

Let us consider a unitary conformal field theory (CFT) and examine the spectrum of scaling dimensions of conformal primary operators.
The scaling dimensions $\Delta_i$ depend on various parameters of CFT such as coupling constants, Casimirs of  internal
symmetry group etc. 
In a close analogy with gauge theories, we assume that CFT depends on a small parameter $1/N$ and the scaling dimensions admit an expansion in powers of $1/N^2$, e.g.
\begin{align}\label{Delta-exp}
\Delta_i =  \Delta_i^{(0)} + {1\over N^2}  \Delta_i^{(1)}+ {1\over N^4}  \Delta_i^{(2)} + \cdots\,.
\end{align} 
Here the leading term $\Delta_i^{(0)}$ is  an eigenvalue  of the dilatation operator in the planar limit, 
the subleading terms can be obtained by diagonalizing the dilatation operator perturbatively in powers of $1/N$. 

In this note we ask the question what happens when the scaling dimensions of two operators collide in the planar limit 
for some value of the coupling constant, $\Delta_1^{(0)} =\Delta_2^{(0)}$. This occurs in particular in maximally supersymmetric 
$\mathcal N=4$ super Yang-Mills theory (SYM) for Konishi and double-trace operators  (defined in \re{Born} below).
The scaling dimension of the double-trace operator is protected from corrections in the planar limit, whereas the scaling dimension of 
the Konishi operator is an
increasing function of 't Hooft coupling constant that starts at $2+O(\lambda)$ at weak coupling and increases as 
$2 \lambda^{1/4}$ at strong coupling \cite{Gubser:2002tv}. As a consequence, the two levels cross each other at finite
$\lambda$. We shall return to this example below.
 
It is well-known that, in the radial quantization, the dilatation operator  in $D-$dimen\-sional 
Euclidean CFT coincides with the Hamiltonian generating time translations on the cylinder 
$\mathbb R\times S^{D-1}$. In a unitary CFT, it is a hermitian operator with respect to the scalar product given by two-point correlation function of $O_i(x)$ on the cylinder and, therefore, its eigenvalues $\Delta_i$ have to satisfy von Neumann--Wigner 
non-crossing rule \cite{Neumann} stating that the levels of the dilatation operator with the same symmetry cannot cross. 

How can we reconcile the non-crossing rule with the fact that the functions $\Delta_i^{(0)}$ providing the
leading (planar) correction to \re{Delta-exp} can cross each other? The situation here is very similar to
that in quantum mechanics --  
the perturbative expansion \re{Delta-exp} is well-defined
if the  energy level separation $\Delta_1 - \Delta_2$ is much larger compared to their `interaction energy'
defined by the leading nonplanar $O(1/N)$ 
correction to the dilatation operator. In the opposite limit,
for 
\begin{align}\label{crossing}
\var\equiv  \Delta_1 - \Delta_2 = O(1/N)\,,
\end{align}
the perturbative expansion \re{Delta-exp} becomes singular due to small denominators and needs to be resummed to all orders in $1/N$.
As we show below, the resummed expressions for $\Delta_i$ indeed respect the non-crossing rule.

\section{Resummed scaling dimensions}

Let us consider a pair of conformal operators $O_1$ and $O_2$ whose scaling dimensions depend on the coupling constant and
admit the $1/N$ expansion \re{Delta-exp}. For the sake of simplicity we take $O_i(x)$ to be scalar operators normalized in such a way 
that their two-point functions are given by
\begin{align}\label{OO}
\vev{O_i(x) O_j(0)} =  {\delta_{ij}\over (x^2)^{\Delta_i}}\,.
\end{align}
We assume that for $N\to \infty$ their scaling dimensions intersect at some value
of the coupling constant, $\Delta_1^{(0)} =\Delta_2^{(0)}$, and remain separated
from the rest of the spectrum of $\Delta_i$'s by a gap that remains finite for $N\to\infty$. We show in this section that for 
$N$ large but finite, the scaling dimensions $\Delta_1$ and $\Delta_2$ do not intersect as one varies the coupling constant,
the minimal distance between the two curves scales as $O(1/N)$.
\footnote{A similar result was independently obtained by Fernando Alday (private communication).}

Let $O_1$ and $O_2$ be the eigenstates of the dilatation operator in the planar limit.
The conformal symmetry implies that away from the crossing point, for $\Delta_1^{(0)} \neq \Delta_2^{(0)}$, the two-point correlation 
function $\vev{O_1(x)O_2(0)}$ should vanish for $N\to\infty$. This does not mean however that $\vev{O_1(x)O_2(0)}$
can not be different from zero. For $N$ large but finite it can receive corrections suppressed by powers of $1/N$ 
\begin{align}\label{off}
\vev{O_1(x)O_2(0)} = {1\over N} \varphi(x^2)  + \dots  \,,
\end{align} 
where dots denote subleading corrections. In the similar manner, the leading nonplanar corrections 
to the diagonal correlation functions $\vev{O_i(x)O_i(0)}$ scale as $O(1/N^2)$.

The function $\varphi(x^2)$ depends on the coupling constant and describes the mixing between the operators $O_1$ and $O_2$.
If the scaling dimensions of the operators $O_1$ and $O_2$ satisfy  \re{crossing} and are separated from the rest of the spectrum
by a finite gap, we expect that the conformal operators are given by their linear combinations  
\begin{align}\label{Opm}
O_+ = O_1 + c_2 O_2\,,\qqqquad O_- = O_2 + c_1 O_1\,.
\end{align}
Requiring $\vev{O_+(x) O_-(0)}=0$ we find with a help of \re{Opm}, \re{OO} and \re{off} that, in the vicinity of the crossing point,
the conformal symmetry restricts the possible 
form of the function $\varphi(x^2)$ 
\begin{align}\notag\label{match}
{1\over N} \varphi(x^2)(1+c_1 c_2) {}& = -c _1 (x^2)^{- \Delta_1}-c _2 (x^2)^{- \Delta_2 }
\\
{}&=-(x^2)^{- \Delta_1 }\big[c_1+c_2 + c_2\, \var \ln x^2\big]\,.
\end{align}
Here in the second relation we took into account \re{crossing} and neglected terms proportional to $\epsilon^2\sim 1/N^2$. Then, going to $N\to\infty$ limit on the both sides of \re{match}  and taking
into account that $\var = O(1/N)$, we find $c_2=-c_1+O(1/N)$ together with
\begin{align}\label{phi}
\varphi(x^2) = {\gamma  \ln x^2 \over (x^2)^{\Delta_1 }}\,,  
\end{align} 
where $\gamma$ does not depend on $N$ and satisfies 
\begin{align} \label{eq}
{}& c_1^2  + c_1 {\var N \over \gamma} - 1=0\,.
\end{align}
We can use this equation to express the coefficients $c_1$ and $c_2$ in terms of $\gamma$ and  obtain from \re{Opm} the conformal operators $O_\pm$ that take into account the leading nonplanar correction.~\footnote{The two solutions to \re{eq} give rise to the same operators $O_\pm$ (up to an overall normalization factor).}

To determine the scaling dimensions of the  operators \re{Opm}, we examine their two-point correlation functions 
\begin{align}\notag\label{O-norm}
\vev{O_-(x) O_-(0)} {}&= {1\over (x^2)^{\Delta_2 }} +{c_1^2\over (x^2)^{\Delta_1 }} + {2c_1 /N\over (x^2)^{\Delta_1 }}   \gamma \ln x^2  
\\
{}&={1+c_1^2  \over (x^2)^{\Delta_2}}\bigg[1+ { c_1(2 \gamma/N-c_1\, \var)  \over 1+c_1^2}  \ln x^2\bigg]
\end{align}
and similar for $ \vev{O_+(x) O_+(0)}$. Matching this relation into \re{OO} we obtain the leading $O(1/N)$ correction
to the scaling dimension of the operator $O_-$  
\begin{align}\notag
\Delta_- {}& = \Delta_2 - {c_1(2  \gamma/N-c_1\, \var )\over 1+c_1^2} 
  =\Delta_2- {\gamma\over N}c_1 \,,
\end{align} 
where in the last relation we applied \re{eq}. Going through similar calculation of $\Delta_+$ we find
\begin{align} \label{fin}
\Delta_\pm  {}& ={\Delta_1+\Delta_2\over 2} \pm \sqrt{{ \var^2\over 4}+{\gamma^2\over N^2}}  \,,
\end{align} 
with $\var=\Delta_1 - \Delta_2$. We recall that this relation holds up to corrections suppressed by powers of $1/N$.  

The following comments are in order.

An attentive reader will likely notice that $\Delta_\pm$ coincide with energies of a two-level system  
with a Hamiltonian
\begin{align}
\mathbb H = \left[\begin{array}{cc} \Delta_1 & \gamma/N \\ \gamma/N & \Delta_2 \end{array}\right]\,.
\end{align} 
Indeed, we can use \re{OO}, \re{off} and \re{phi} to verify that this matrix defines the action of
the dilatation operator on the operators $O_1$ and $O_2$, that is $i [\mathbb{D},O_i(0)]= \mathbb{H}_{ij} \, O_j(0)$. 
Then, the construction of the conformal operators \re{Opm} follows the usual consideration of the operator mixing in 
gauge theory at one loop with the only difference that $1/N$ plays the role of 't Hooft coupling constant.~\footnote{I would like to thank
Sergey Frolov for suggesting this interpretation.}

It is instructive to compare \re{fin} with the general expression \re{Delta-exp}. Expanding $\Delta_\pm$ in powers of $1/N$ we find from \re{fin}
\begin{align}\notag\label{div}
\Delta_+ =  \Delta_1 + {\gamma^2/ \var\over N^2} { } -  {\gamma^4/\var^3\over N^4} + \cdots\,,
\\
\Delta_- =  \Delta_2 - {\gamma^2/ \var\over N^2} { } +  {\gamma^4/\var^3\over N^4} + \cdots\,.
\end{align}
Comparison with \re{Delta-exp} shows that $\Delta_\pm^{(p)}\sim \gamma^{2p}/\var^{2p-1}$ and, therefore, the expansion  \re{Delta-exp}
becomes singular for $\var=O(1/N)$, in agreement with our expectations. The relation \re{fin} takes into account an infinite
class of the leading corrections $O\left(1/(\var^{2p-1} N^{2p})\right)$ to all orders in $1/N$ and it remains finite for $\epsilon\to 0$. At the
crossing point, for $\var=0$, we find
from \re{fin} that $\Delta_\pm = \Delta_1\pm \gamma/N$, so that the leading nonplanar correction to the resummed scaling dimensions scales as $O(1/N)$ 
(and not  $O(1/N^2)$ as one would expect from \re{Delta-exp}).
\footnote{Analogous phenomenon has been previously observed for critical dimensions of composite operators 
in the nonlinear $O(N)$ sigma-model \cite{Derkachov:1998js}. At large $N$ their expansion runs in powers of $1/N$ 
but the leading nonplanar correction scales as $O(1/\sqrt{N})$ due to the level crossing.
}
 
\begin{figure}[t!bp]
\psfrag{g}[cc][cc]{$g$}
\parbox[c]{\textwidth}{
\includegraphics[height = 0.22\textheight]{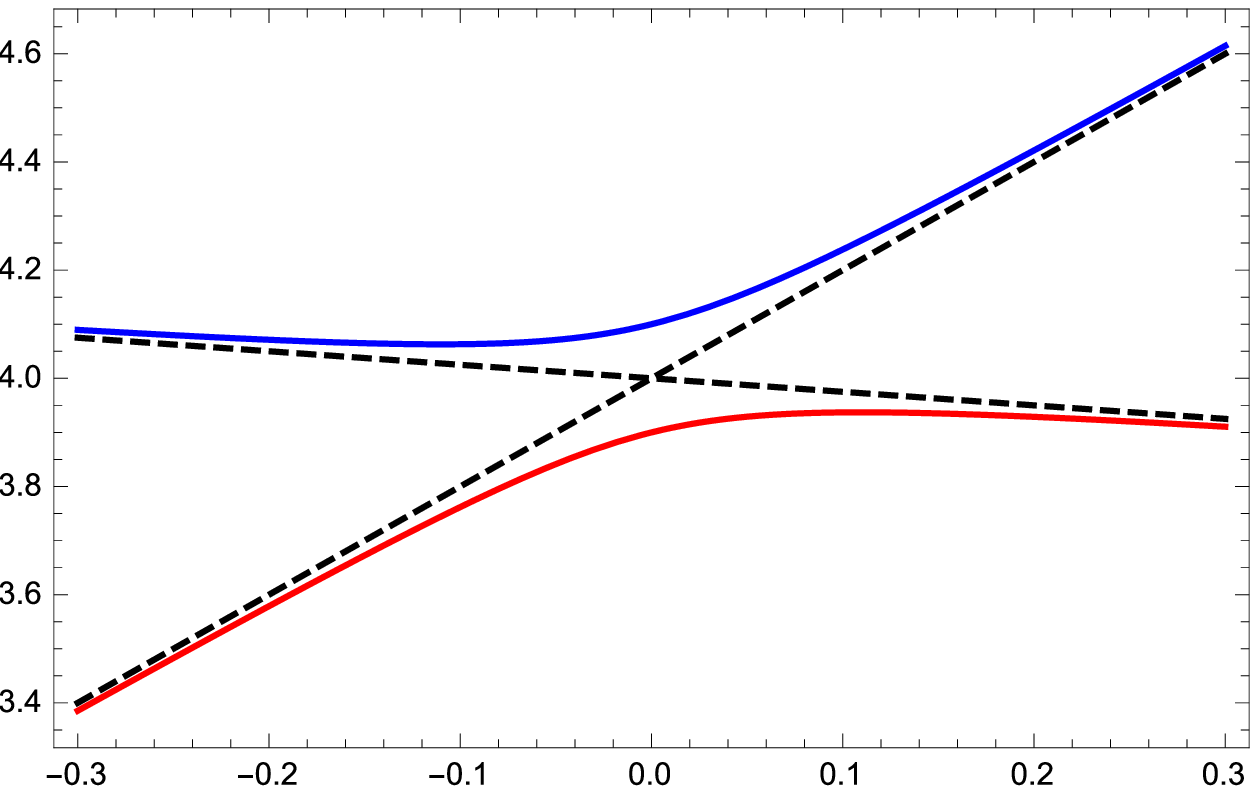}\qquad\includegraphics[height = 0.22\textheight]{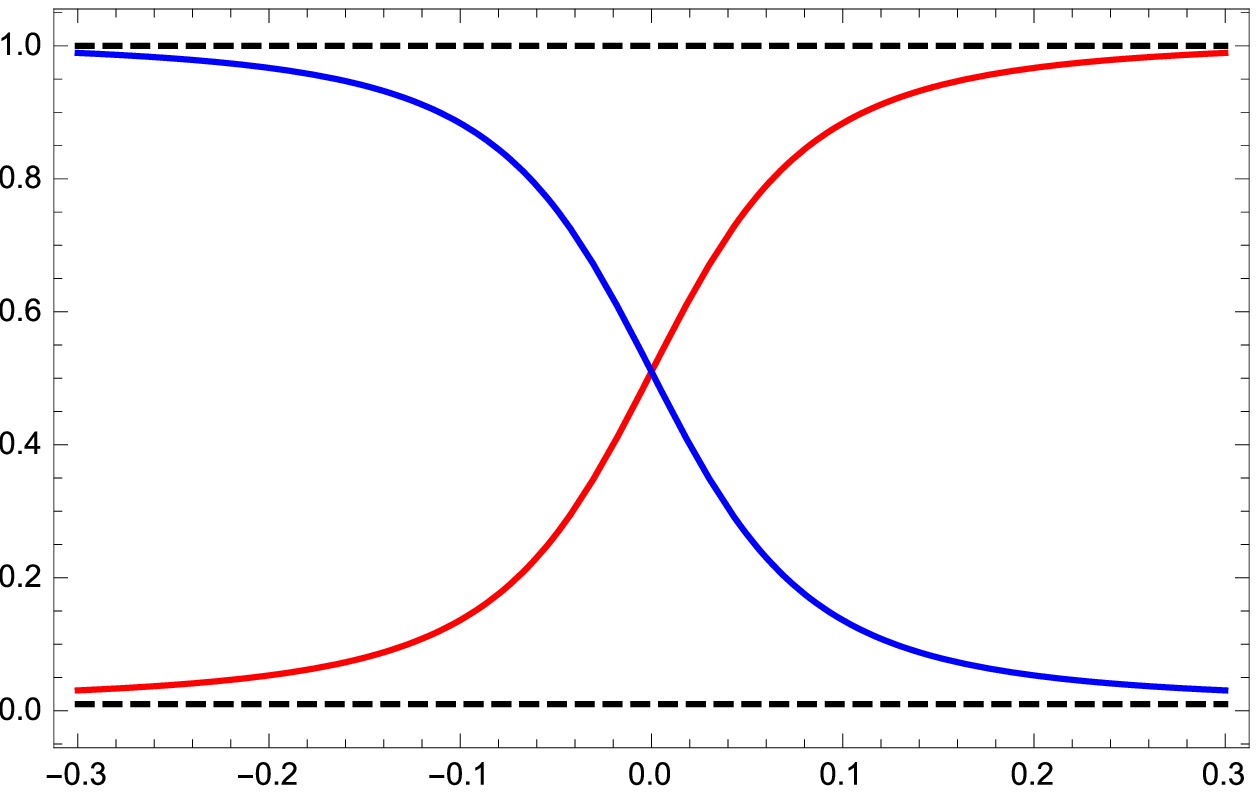}}
\caption{The scaling dimensions  $\Delta_\pm$ (left panel) and the OPE coefficients $(C_{\phi\phi O_\pm} / C_{\phi\phi O_1} )^2$   (right panel) in the transition region \re{crossing} as a function of the coupling constant $g$ for $\gamma/N=0.1$. The dashed lines represent the same quantities in the planar
limit, $\Delta_i(g) = \alpha_i g + O(g^2)$.}
\label{fig}
\end{figure}

Let us assume that the scaling dimensions $\Delta_1$ and $ \Delta_2$ are continuous functions of the coupling
constant $g$, intersecting in the planar limit at $g=0$. Then, the relation \re{fin} defines two functions $\Delta_\pm(g)$ that satisfy the non-crossing rule. Namely, the two levels $\Delta_+(g)$ and
$\Delta_-(g)$  approach each other for $g=0$ but remain separated by a finite gap  (see Figure~\ref{fig})
\begin{align}
|\Delta_+-\Delta_-| \, \ge \, {2\gamma\over N}\,.
\end{align}  
We recall that the parameter $\gamma$ defines the leading nonplanar correction to the two-point correlation function \re{off} and \re{phi}.
As can be seen from Fig.~\ref{fig}, the level crossing leads to a change of behavior of the scaling dimensions in the
vicinity of the crossing point. Away from the crossing point, the resummed scaling dimensions  approach their values
in the planar limit, e.g. $\Delta_+ \approx \Delta_1$ at large negative $g$ goes into $\Delta_+\approx \Delta_2$ at large positive $g$.

\section{Resummed OPE coefficients}

In the previous section, we defined the conformal primary operators $O_\pm$ in the transition region \re{crossing} 
and determined their scaling dimensions \re{fin}. Let us examine the properties of the OPE coefficients $C_{\phi\phi O_\pm}$
defining the correlation function $\vev{ \phi \phi  O_{\pm}}$ involving scalar conformal operator $\phi$  
with the scaling dimension $\Delta_{\phi}$ separated from $\Delta_\pm$ by a finite gap, $|\Delta_{\phi} - \Delta_\pm| = O(N^0)$.

Let us consider a four-point function of the operators $\phi$ and decompose it over the conformal blocks  describing the
contribution of different conformal multiplets
\begin{align}\label{4pt}
\vev{\phi(x_1)\phi(x_2)\phi(x_3)\phi(x_4)} = {1\over (x_{12}^2 x_{34}^2)^{\Delta_\phi}}\left[\mathcal  F_ {O_+}(u,v)+\mathcal  F_{O_-}(u,v)+\sum_{\phi_i} \mathcal F_{\phi_i}(u,v)\right],
\end{align}
where $x_{ij}=x_i-x_j$, $u=x_{12}^2 x_{34}^2/(x_{13}^2 x_{24}^2)$ and $v=x_{23}^2 x_{14}^2/(x_{13}^2 x_{24}^2)$. For $x_{12} \to 0$, or equivalently
$u\to 0$ and $v\to 1$, the contribution of the operators $O_\pm$ to \re{4pt} takes the form 
\begin{align}\label{F-as}
\mathcal  F_ {O_\pm}(u,v)= C_{\phi\phi O_\pm}^2 u^{\Delta_\pm/2} + \dots\,,
\end{align}
where dots denote terms suppressed by powers of $u$ and $1-v$.

The OPE coefficients in \re{F-as} are given by $C_{\phi\phi O_\pm}^2\sim {\vev{\phi\phi O_\pm} \vev{O_\pm \phi\phi}/ \vev{O_\pm O_\pm}}$.
Replacing $O_\pm$ with their explicit expressions \re{Opm}, we find using \re{eq}
\begin{align}\notag\label{C2}
C^2_{\phi\phi O_+} = {\left(C_{\phi\phi O_1}-c_1  C_{\phi\phi O_2}\right)^2 \over 1+c_1^2}\,,
\\
C^2_{\phi\phi O_-} = {\left(C_{\phi\phi O_2}+c_1  C_{\phi\phi O_1}\right)^2 \over 1+c_1^2}\,,
\end{align}
where $C_{\phi\phi O_i}$ are the OPE coefficients for the operators $O_i$ in the planar limit and
\begin{align}\label{c1}
c_1 =  -{N\epsilon\over 2\gamma}+ \sqrt{{N^2\epsilon^2\over 4\gamma^2}+1}= {\gamma\over N\epsilon} - \lr{\gamma\over N\epsilon}^3
+\dots\,,
\end{align}
for $\epsilon>0$.
The relation \re{C2} defines two smooth functions of $\epsilon$ that cross each other for $\epsilon\neq 0$ and take the 
values  $C^2_{\phi\phi O_\pm} =  (C_{\phi\phi O_1} \pm C_{\phi\phi O_2})^2/2 $ at the crossing point  $\epsilon=0$.

The four-point correlation function \re{4pt} is a well-defined function of the parameters of the underlying CFT. In particular, at large $N$ it admits an expansion that runs in powers of $1/N^2$. It is natural to require that each term of this expansion should be finite for 
$\epsilon\to 0$. As we will see in the moment, this requirement leads to nontrivial consequences for the OPE coefficients 
$C_{\phi\phi O_i}$.

The conformal blocks entering the right-hand side of \re{4pt} depend on the scaling dimensions of the operators $\phi_i$. Since
$\Delta_{\phi_i}$ are well-separated from each other in the transition region \re{crossing}, the corresponding conformal
blocks $\mathcal F_{\phi_i}$ automatically satisfy the regularity condition mentioned above.
This is not the case however for the conformal blocks $\mathcal  F_ {O_\pm}$. 
According to \re{div} and \re{c1}, the large $N$
expansion of $\Delta_\pm$ and $c_1$ is singular in the region \re{crossing} and, as a consequence, the coefficients of the expansion of \re{F-as} in powers of $1/N$ contain poles in $\epsilon$. They cancel however in the sum  $\mathcal  F_ {O_+}+\mathcal  F_{O_-}$ that remains finite for $\var\to 0$. Namely,
\begin{align}\label{odd}
\mathcal  F_ {O_+}+\mathcal  F_{O_-}  =\lr{ C^2_{\phi\phi O_1} u^{\epsilon/4}+C^2_{\phi\phi O_2} u^{-\epsilon/4}}
- 2\gamma C_{\phi\phi O_1} C_{\phi\phi O_2} { u^{\epsilon/4}-u^{-\epsilon/4} \over N \epsilon} + O(1/N^2)\,.
\end{align}
Notice that the second term on the right-hand side is finite for $\epsilon\to 0$ but it scales as $O(1/N)$. 

It is straightforward to verify
that the subleading corrections to \re{odd} involve both even and odd powers of $1/N$. This seems to be in contradiction with the
requirement that large $N$ expansion of the correlation function \re{4pt} should run in even powers of $1/N$. We can eliminate all
terms with odd powers of $1/N$ by imposing the additional condition on the OPE coefficients \footnote{There is obviously 
another possibility  $ C_{\phi\phi O_1} /C_{\phi\phi O_2} =O(1/N)$ but it leads to the same result for $C_{\phi\phi O_\pm}^2$. }
\begin{align}\label{sol2}
  {C_{\phi\phi O_2}\over C_{\phi\phi O_1}} =O(1/N)\,.
\end{align}
This relation implies that if the scaling dimensions of two operators $O_1$ and $O_2$ approach 
each other and satisfy \re{crossing}, their OPE coefficients in the planar limit have to differ by a large factor of $N$. 

Combining together \re{sol2} and \re{C2}, we obtain the following expression for
the resummed OPE coefficients
\begin{align}\label{C-dif1}
C_{\phi\phi O_\pm}^2  
 = \frac12 C_{\phi\phi O_1}^2  \bigg[ 1\pm  {\var \over  \sqrt{\var^2+4\gamma^2/  N^2 }}\bigg] \,,
\end{align}
where $\var=\Delta_1 - \Delta_2$ and $C_{\phi\phi O_1}$  are the OPE coefficients in the planar limit.
Let us examine the expansion of \re{C-dif1} at large $N$ and $\epsilon>0$
\begin{align}\notag\label{C-pole}
{}& C_{\phi\phi O_+}/C_{\phi\phi O_1}=  1 - {\gamma^2 \over 2 N^2 \var^2} + O(1/N^4) \,,
\\
{}& C_{\phi\phi O_-}/C_{\phi\phi O_1}=  {\gamma \over N \var}\left(1 - {3\gamma^2 \over 2N^2 \var^2} + O(1/N^4) \right)\,.
\end{align}
Notice that the leading correction to $C_{\phi\phi O_-}$ contains a pole at $\epsilon=0$. Its appearance is an arfifact of the large $N$
expansion. As follows from \re{C-dif1}, the resummed expression for $C_{\phi\phi O_-}$ is finite and it takes the value 
$C_{\phi\phi O_1}/\sqrt{2}$ for $\epsilon=0$.  Away from the crossing point, for $|\epsilon N|\gg 1$,
the structure constants $C_{\phi\phi O_\pm}$ coincide with those in the planar limit, $C_{\phi\phi O_1}$ and
$C_{\phi\phi O_2}$.

The dependence
of the  OPE coefficients $(C_{\phi\phi O_\pm} / C_{\phi\phi O_1} )^2$ on the coupling constant in the vicinity of the crossing point is shown in Fig.~\ref{fig}.
We demonstrate in the next section that the relations \re{C-dif1} and \re{C-pole} are in a good agreement 
with the known properties of  three- and four-dimensional CFT's.

We recall that the relation \re{C-dif1} was obtained from the requirement for the four-point correlation function \re{4pt} to have a regular
$1/N^2$ expansion as $\epsilon\to 0$. We substitute \re{C-dif1} into \re{odd} and verify that this condition is indeed verified
\begin{align} \label{sum-F}
\mathcal  F_ {O_+}+\mathcal  F_{O_-} {}& \sim u^{\Delta_1 /2+\epsilon/4} 
\left\{ 1
+  {\gamma^2\over 8N^2}\left[(\ln u)^2 -{\epsilon\over 6} (\ln u)^3+O(\epsilon^2)\right] + O(1/N^4)\right\}.
\end{align}
A distinguished feature of this relation as compared with the analogous asymptotic behaviour of the conformal blocks 
is the appearance of terms enhanced by powers of $\ln u$. 
For $\epsilon = 0$ the $1/N-$expansion on the right-hand side of \re{sum-F} can be resummed leading to
\begin{align}
\mathcal  F_ {O_+}+\mathcal  F_{O_-} {}& \sim
 \frac12  u^{( \Delta_1+\gamma/N)/2} +  \frac12   u^{( \Delta_1-\gamma/N)/2} \,.
\end{align}
We recognize the two terms on the right-hand side as describing the leading $u\to 0$ asymptotics of conformal blocks corresponding
to two conformal primary operators with the scaling dimensions $ \Delta_1\pm \gamma/N$.
 
\section{Examples of level-crossing }

In this section, we review the known examples of the level-crossing phenomenon in three- and four-dimensional CFT's and make 
a comparison with the results obtained in the previous sections. 

\subsection{Level-crossing in $\mathcal N=4$ SYM}\label{N4}

Our first example concerns the mixing of the Konishi and double-trace operators in four-dimensional  
$\mathcal N=4$ SYM theory. As was mentioned in the Introduction, the scaling dimensions of these operators collide in the 
planar limit at finite value of the coupling constant. 

In planar $\mathcal N=4$ SYM, the Konishi and double-trace operators take the form
\begin{align}\notag\label{Born}
{}& \mathcal O_K ={1\over  N}  \tr[\Phi^I \Phi^I]\,, 
\\
{}& \mathcal O_{D}={1\over N^J} \tr[\Phi^{(I_1} \dots \Phi^{I_J)}]\tr[\Phi^{(I_1} \dots \Phi^{I_J)}]\,,
\end{align}
where $\Phi^I$ (with $I=1,\dots,6$) are real scalar fields and $\tr[\Phi^{(I_1} \dots \Phi^{I_J)}]$ is a symmetric traceless $SO(6)$  
tensor. The normalization factors on the right-hand side of \re{Born} were introduced for the two-point correlation functions of the
operators to scale as $O(N^0)$ at large $N$.

At Born level, for zero coupling constant, the scaling dimensions of the operators \re{Born} are $\Delta_K =2$ and $\Delta_D=2J$,
respectively. At strong coupling, the scaling dimensions can be computed using the AdS/CFT correspondence, see e.g. \cite{Dolan:2001tt,Gromov:2009zb}
\begin{align}\notag
{}& \Delta_D = 2J- {2(J-1)J(J+2)\over N^2}+\dots\,,  
\\
{}& \Delta_K =  2 \lambda^{1/4} -2 + {2\over \lambda^{1/4}} + \dots \,,
\end{align}
where dots denote terms suppressed by powers of $1/N^2$ and $\lambda^{-1/2}$. We observe that,  at strong coupling, for $J\approx \lambda^{1/4}$ the two
levels cross each other. 

To establish the connection with the results of the previous sections, we examine three-point correlation functions 
of the operators \re{Born} and half-BPS operators  
\begin{align}\label{1/2}
O_J(x,Y) = {1\over N^{J/2}} Y_{I_1}\dots Y_{I_J}\tr[\Phi^{(I_1} \dots \Phi^{I_J)}]\,,
\end{align}
where $Y_I$ is an auxiliary six-dimensional null vector, $Y_I^2=0$. This operator carries the $R-$charge $J$ and it 
scaling dimension is protected from quantum corrections, $\Delta_{O_J} = J$. The form of the three-point functions 
is fixed by conformal symmetry
\begin{align}\notag\label{three}
\vev{O_J(1) O_J(2) O_K(0)} = (Y_1 Y_2)^J {C_{JJK}\over x_1^{\Delta_K} x_2^{\Delta_K} x_{12}^{2J-\Delta_K}}\,,
\\
\vev{O_J(1) O_J(2) O_D(0)} = (Y_1 Y_2)^J {C_{JJD}\over x_1^{\Delta_D} x_2^{\Delta_D} x_{12}^{2J-\Delta_D}}\,.
\end{align}
where $O_J(i) \equiv O_J(x_i,Y_i)$.
The correlation function in the second line of \re{three} factors out in the planar limit into the product of two-point
correlation functions of half-BPS operators \re{1/2} leading to  
\begin{align}\label{C-D}
C_{JJD}= 1 + O(1/N^2)\,.
\end{align}
The OPE coefficient $C_{JJK}$ was computed at strong coupling in Ref.~\cite{Minahan:2014usa}. It was found that $C_{JJK}$  
develops a pole at $J=\Delta_K/2\approx \lambda^{1/4}$ ~\footnote{The three-point correlation functions in which the dimension 
of one operator equal to  the sum of the other two are known as extremal correlators. The appearance of poles 
is a generic feature of the extremal correlators in the AdS/CFT correspondence \cite{Freedman:1998tz,D'Hoker:1999ea,Liu:1999kg}. \label{ft}} 
\begin{align}\label{C-K}
C_{JJK} = {J^{3/2}\sqrt{M} \over 2N(J-\lambda^{1/4})}\,,
\end{align}
where $M=(J+1)(J+2)^2(J+3)/12\approx J^4/12$ is the dimension of the $J-$symmetric traceless representation of the
$SO(6)$. As was argued in \cite{Minahan:2014usa}, the appearance of a pole in \re{C-K} leads to the mixing of the Konishi operator $O_K$ and the
double-trace operator $O_D$.

Let us compare the relations \re{C-D} and \re{C-K} with the general expression \re{C-pole}. Upon identification of the operators, $\phi_i = O_J$, $O_+ =O_D$ and $O_-=O_K$, we find a perfect agreement with the leading term in \re{C-pole} for 
\begin{align}
\epsilon=\Delta_D-\Delta_K\approx 2(J-\lambda^{1/4})\,,\qqqquad 
\gamma=\lambda^{3/8}\sqrt{M}\approx {J^{7/2} \over 2\sqrt{3}}\,.
\end{align}
Substituting these expression into \re{C-pole}, we can determine the subleading terms in \re{C-D} and \re{C-K}. Moreover, 
we can apply \re{fin} and \re{C-dif1} to obtain the resummed expressions for the scaling dimensions and the OPE coefficients in the
transition region $|J-\lambda^{1/4}|=O(1/N)$
\begin{align} \notag\label{C-finite}
{}& \Delta_\pm   =J+\lambda^{1/4} \pm \sqrt{(J-\lambda^{1/4})^2+{J^7\over 12 N^2}}  \,,
\\
{}& C_{JJ O_\pm}^2   = \frac12  \bigg[ 1\pm  {J-\lambda^{1/4} \over  \sqrt{(J-\lambda^{1/4})^2+{J^7/(12 N^2)} }}\bigg] \,,
\end{align}  
where the subscripts $+/-$ correspond to the double-trace and Konishi operators, respectively. At the crossing point, for 
$J\approx\lambda^{1/4}$, the level splitting is
\begin{align}
|\Delta_+-\Delta_-| = {J^{7/2}\over N\sqrt{3}}\,.
\end{align}
This expression is in agreement with the one given in Ref.~\cite{Minahan:2014usa} (see the latest version). The structure
constants \re{C-finite} are finite in the transition region and do not require the additional renormalization
of operators advocated in \cite{Minahan:2014usa}.

The dependence of $\Delta_\pm$ on the coupling constant in the vicinity of  $ \lambda^{1/4}\approx J$ follows the same pattern as
shown in Fig.~\ref{fig}. In particular, examining $\Delta_-$ as a function of the coupling constant, we find that it grows for $\lambda^{1/4}\ll J$ 
as the scaling dimension of the Konishi operator $\Delta_K$ and, then, approaches the scaling dimension of the double-trace operator
$\Delta_D$ for $\lambda^{1/4}\gg J$.~\footnote{To the left of the crossing point, for $\lambda^{1/4} \ll J$, we have $O_+ =O_D$ and $O_-=O_K$, whereas
to the right of the crossing point, for $\lambda^{1/4}\gg J$, the operators are exchanged $O_+ =O_K$ and $O_-=O_D$. At the crossing
point, for $\lambda^{1/4}\approx J$, the operators are maximally mixed, $O_\pm = O_K\pm O_D$. }
 As a consequence, it satisfies the relation $\Delta_- < \Delta_D< 2J$, so that the correlation function $\vev{O_J(1) O_J(2) O_-(0)}$ can not be extremal for finite $N$ (see footnote \ref{ft}).~\footnote{The extremal situation can arise however 
for the correlation functions of protected operators $\vev{O_{J_1}(1) O_{J_2}(2) O_{J_3}(3)}$ for $J_3=J_1+J_2$. In this case, the level crossing does not occur since the scaling dimensions and the OPE coefficients
are protected by $\mathcal N=4$ superconformal symmetry. The calculation of such correlators in the AdS/CFT turns out to be a nontrivial task \cite{D'Hoker:1999ea,Liu:1999kg,Arutyunov:1999en,Arutyunov:2000ima}. }

As the second example, we consider the mixing of leading-twist operators carrying a nonzero Lorentz spin $S$. The
scaling dimension of twist-two operators grows at large spin as 
$\Delta_2=2+S+ \gamma_S$, where
$\gamma_S=2 \Gamma_{\rm cusp}(\lambda) \ln  \bar S + O(S^0)$ 
with 
$\Gamma_{\rm cusp}(\lambda)$ being the cusp anomalous dimension and $\bar S = S \e^{ \gamma_{\rm E}}$ (with $\gamma_{\rm E}$ being the Euler constant)  \cite{Korchemsky:1988si}. 
For twist-four operators, the lowest scaling dimension scales at large spin as $\Delta_4=4+S - c(N)/S^2$ with positive $c(N)$ \cite{Alday:2007mf,Fitzpatrick:2012yx,Komargodski:2012ek}. Comparing $\Delta_2(S)$ and $\Delta_4(S)$ we find that the two functions cross each other at $\gamma_S=2 + O(1/S^2)$, or
equivalently $\Gamma_{\rm cusp}(\lambda)\sim 1/\ln \bar S$. 
The important difference with the previous example is that the crossing point is located
at weak coupling and, therefore, it can be analyzed using perturbation theory. 
 
The three-point correlation function
of two half-BPS operators \re{1/2} with charge $J=2$ and twist-two operator with large spin $S$ has been computed at weak coupling in Ref.~\cite{Alday:2013cwa} leading to the following result for 
the corresponding OPE coefficient 
\begin{align}\label{tw2}
C_2(S)/C_2^{(0)}(S) ={1\over N}  \Gamma\left(1-\ft12\gamma_S(\lambda)\right) \e^{-\frac12 \gamma_S(\lambda)[ \gamma_{\rm E}+ \alpha(\lambda)] + \beta(\lambda)},
\end{align}
where $C_2^{(0)}(S)=[2 \Gamma^2(S+\gamma_S/2+1)/\Gamma(2S+\gamma_S+1)]^{1/2}$ and functions $\alpha(\lambda)$ and $\beta(\lambda)$ are known explicitly up to two loops. 
The expression on the right-hand side of \re{tw2} was obtained for $S\gg 1$ with $\lambda\ln S= \text{fixed}$, it
takes into account all corrections of the form $\lambda^k (\ln S)^n$ (with $1\le n \le k$).
For twist-four operators of the schematic form $O_2 \partial^S O_2$ the analogous OPE coefficient is
\begin{align}\label{tw4}
C_4(S) /C_4^{(0)}(S) =1 + O(1/N^2)\,,
\end{align}
where $C_4^{(0)}(S)= [(S+1)!(S+2)!/(2S+1)!]^{1/2}$.

We observe that the twist-two OPE coefficient \re{tw2} develops a pole at $\gamma_S(\lambda)=2$.  
Since $\Gamma_{\rm cusp}=\lambda/(4\pi^2) +O(\lambda^2)$ at weak coupling, the corresponding
value of the spin $S$ is exponentially large,  
$
S=\exp\lr{{4\pi^2/ \lambda} - \gamma_{\rm E}+ O(\lambda)}
$.
Denoting $\epsilon =  \Delta_4-\Delta_2=2-\gamma_S$, we find that for $\epsilon\to 0$ the relations
\re{tw2} and \re{tw4} are in a perfect agreement with \re{C-pole}
\begin{align}
{C_2(S)\over C_4(S)} ={2 \e^{- \gamma_{\rm E}}\over  \epsilon\, N S } \left[1+ O(\lambda,1/N^2)\right]\,.
\end{align}
The residue at $1/\epsilon$ pole defines the interaction energy 
$\gamma$ between twist-two and twist-four operators
\begin{align}
\gamma={2 \e^{- \gamma_{\rm E}}\over S}=2 \e^{-{4\pi^2/\lambda}+ O(\lambda)}\,,
\end{align}
which is exponentially small at weak coupling.
Applying \re{fin} and  \re{C-dif1}, we can obtain the resummed expressions for
the scaling dimensions of these operators and their OPE coefficients in the transition region $|\Delta_4-\Delta_2|= O(1/N)$.
 
\subsection{Level-crossing in three-dimensional CFT}

Recently an important progress has been achieved in understanding the properties of three-dimensional unitary CFT's within 
the conformal bootstrap approach. It was found that the constraints of crossing symmetry and unitarity restrict the
possible values of the scaling dimensions of the operators $\Delta_i$ in a highly nontrivial way \cite{ElShowk:2012ht,El-Showk:2014dwa}.

For our purposes we shall examine the extremal solutions to these constraints living on the boundary of the allowed region of $\Delta_i$'s. 
The spectrum of the corresponding three-dimensional unitary CFT's can be uniquely reconstructed and it is parameterized by the dimension  
$\Delta_\sigma$ of the leading $\mathbb Z_2-$odd scalar operator $\sigma$. As was shown in \cite{El-Showk:2014dwa}, the 
resulting scaling dimensions reveal a dramatic transition at $\Delta_{\sigma}^\star = 0.518154(15)$. The values of $\Delta_i$'s at the
transition point are remarkably close to the expected values of the critical exponents of three-dimensional Ising model.

Another intriguing feature of the 3d Ising point is that some operators disappear from the spectrum. As an example, consider
the leading spin$-2$ operators that appears in the OPE $\sigma\times \sigma$. The lowest lying operator is the stress-energy tensor $T$ whose conformal
weight $\Delta_T=3$ is protected and is independent on $\Delta_{\sigma}$. For the next-to-lowest spin$-2$ operator $T'$, the scaling
dimension and the OPE coefficient at the 3d Ising point are given by $\Delta_{T'}^\star = 5.500(15)$ and 
$(C_{\sigma\sigma T'}^\star)^2 = 2.97(2) \times 10^{-4}$, respectively. However, slightly above this point, for $\Delta_{\sigma} = 0.5182$, 
the OPE coefficient 
decreases  by the order of magnitude $C_{\sigma\sigma T'}^2 =2.21 \times 10^{-5}$ and the scaling dimension changes to 
$\Delta_{T'} =4.334$ (see Figure 16 in Ref.~\cite{El-Showk:2014dwa}).  

\begin{figure}[t!bp]
\hspace*{-7mm} 
\includegraphics[width =1.05\textwidth]{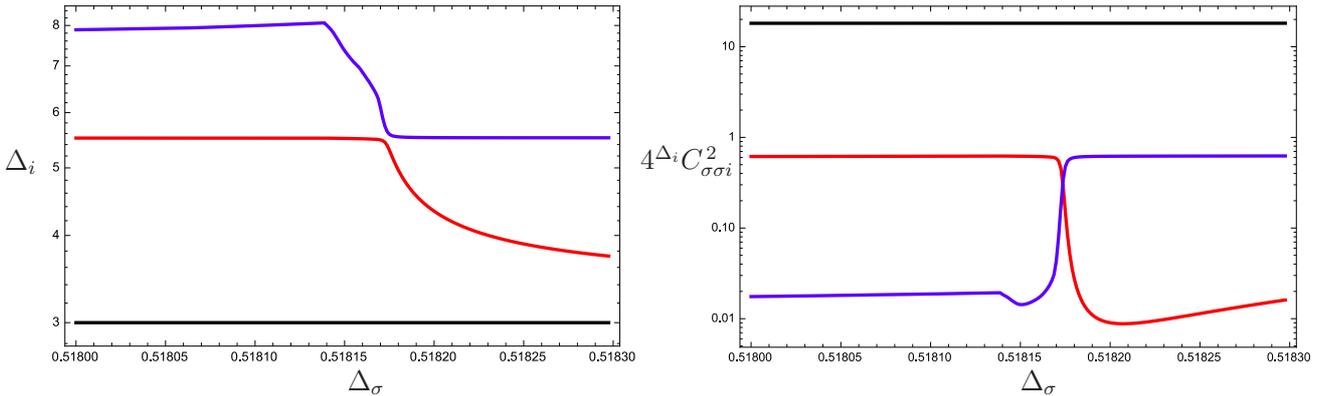} 
\caption{The scaling dimensions of the three lowest spin$-2$ operators and their OPE coefficients
 (multiplied by  $4^{\Delta_i}$) as found in Ref.~\cite{El-Showk:2014dwa}.}
\label{fig3}
\end{figure}   

A close examination of the spectrum of lowest spin $2$ operators (see Figure 15 in Ref.~\cite{El-Showk:2014dwa}) shows that 
in the vicinity of the 3d Ising point the levels approach each other in a pairwise manner. Moreover, the variation of the scaling dimensions
and the OPE coefficients in the transition region follows the same pattern as the one shown in Fig.~\ref{fig}. This suggests that
the spectrum in this region can be described by the relations \re{fin} and \re{C2}. Important difference with the previous case
is that the three-dimensional CFT's under consideration do not have a natural parameter like $1/N$ and the arguments leading to \re{sol2}
and \re{C-dif1} do not apply. In what follows we shall apply  \re{fin} and \re{C2} and 
interpret the parameter $\gamma/N$  as  `interaction energy'  $v$ between levels.

Let us examine the $\Delta_\sigma-$dependence of the scaling dimensions and the OPE coefficients of three lowest levels
shown in Fig.~\ref{fig3}. As was already mentioned, the lowest state with $\Delta_T=3$ (black line) corresponds to the stress-energy tensor.
For the next-to-lowest level (red line), the scaling dimension $\Delta_-$ and the OPE coefficient $C_{\sigma\sigma -}$ vary slowly with
$\Delta_\sigma$ to the left from the 3d Ising point, for $\Delta_\sigma< \Delta_\sigma^*$, and 
approach the following values for $\Delta_\sigma \ll \Delta_\sigma^*$
\begin{align}\label{values}
\Delta_1=5.512\,,\qqqquad  
C_{\sigma\sigma 1}^2= 2.984\times 10^{-4}\,.
\end{align}
For the next-to-next-to-lowest level (blue line),  $\Delta_+$ and $C_{\sigma\sigma +}$  vary slowly with $\Delta_\sigma$  to the right 
from the 3d Ising point, for $\Delta_\sigma >  \Delta_\sigma^*$, and approach the same values \re{values} for $\Delta_\sigma \gg \Delta_\sigma^*$. 

\begin{figure}[t!bp]
\hspace*{-7mm} 
\includegraphics[width =1.05\textwidth]{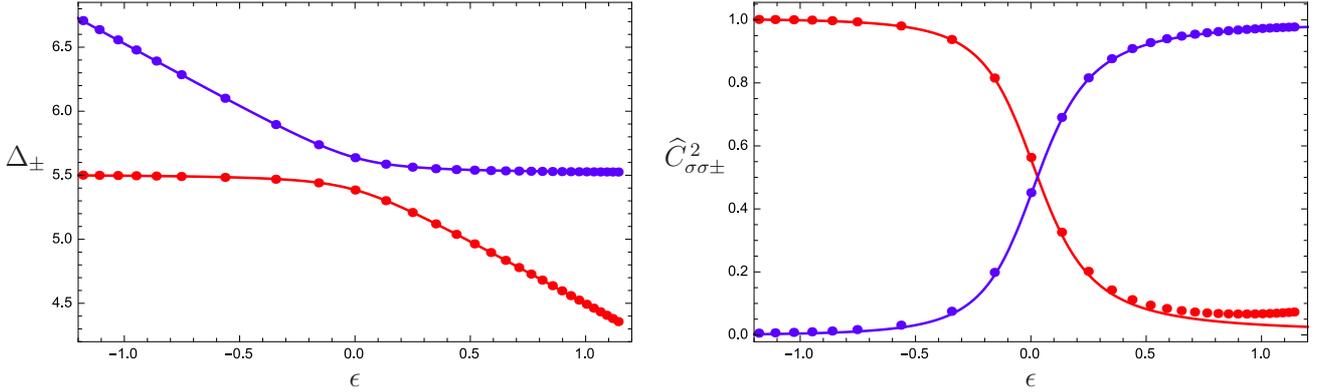} 
\caption{Dependence of the scaling dimensions of the spin$-2$ operators and their OPE coefficients
$\widehat C_{\sigma\sigma \pm}^2 = C_{\sigma\sigma \pm}^2 /C_{\sigma\sigma 1}^2$ 
 on the level splitting $\epsilon$. Dots stand for the exact values found in \cite{El-Showk:2014dwa}, solid lines are described by
 \re{ad-3d} and \re{C-3d}. }
\label{fig2}
\end{figure}

To describe the variation of $\Delta_\pm$ and $C_{\sigma\sigma\pm}$ in the transition region, for 
\begin{align}\label{tr}
 0.51815 <\Delta_\sigma < 0.51820\,,
\end{align}
we introduce the level splitting function $\epsilon=\epsilon(\Delta_\sigma)$ analogous to \re{crossing}
\begin{align}
\epsilon=2\Delta_1-\Delta_- -\Delta_+\,.
\end{align} 
It is straightforward to verify that away from the transition region $\epsilon$ measures the difference 
between the scaling dimensions: $\epsilon=\Delta_--\Delta_+<0$
for $\Delta_\sigma \ll \Delta_\sigma^*$ and  $\epsilon=\Delta_+-\Delta_-<0$ for
$\Delta_\sigma \gg \Delta_\sigma^*$. We find that
$\epsilon$ is a smooth monotonic function of $\Delta_\sigma$ in the transition region \re{tr} and it vanishes for $\Delta_\sigma=0.51817$.
Then, we can invert the function $\epsilon=\epsilon(\Delta_\sigma)$ and obtain the dependence
of $\Delta_\pm$ and $C_{\sigma\sigma \pm}$ on $\epsilon$ as shown in Fig.~\ref{fig2}. 

Applying \re{fin}, we expect that the scaling dimensions $\Delta_\pm(\epsilon)$ in the transition region \re{tr} are given by
\begin{align}\label{ad-3d}
\Delta_\pm(\epsilon) = \Delta_1-{\epsilon\over 2} \pm  \sqrt{{\epsilon^2\over 4} + v^2}\,,
\end{align}
where $\Delta_1$ is defined in \re{values} and $v$ is the interaction energy. For the OPE coefficients, we apply \re{C2} and \re{c1} to get
\begin{align}\notag\label{C-3d}
{}& C^2_{\sigma\sigma +}(\epsilon) = {[C_{\sigma\sigma 1}-c_1  C_{\sigma\sigma 2}]^2 \over 1+ c_1^2 }\,,
\\
{}& C^2_{\sigma\sigma -}(\epsilon) = {[C_{\sigma\sigma 2}+c_1  C_{\sigma\sigma 1}]^2 \over 1+ c_1^2 }\,,
\end{align}
where $c_1(\epsilon)= \lr{\sqrt{\epsilon^2+4v^2}-\epsilon}/(2v)$ and $C_{\sigma\sigma 1}$ is given by \re{values}.
To fix $v$ and $C_{\sigma\sigma 2}$, we compare \re{ad-3d} and \re{C-3d} with the results 
of Ref.~\cite{El-Showk:2014dwa} shown by dots in Fig.~\ref{fig2}. We find a good agreement for
\begin{align}\label{v}
v=0.126  
\,,\qqqquad
C_{\sigma\sigma 2}/C_{\sigma\sigma 1} = 0.057\,.
\end{align}
These values depend on the number of the crossing-symmetry constraints $n_{\rm comp}=231$ used in Ref.~\cite{El-Showk:2014dwa} and they are expected to decrease as $n_{\rm comp}\to\infty$.
An immediate consequence of \re{C-3d} and \re{v} is that the OPE coefficient $C^2_{\sigma\sigma -}(\epsilon)$ decreases by the order of 
magnitude in the transition region \re{tr}. This result is in a quantitative agreement with the expected properties of the operator $T'$
mentioned in the beginning of the subsection.

We observe from Fig.~\ref{fig2} that the OPE coefficients $C^2_{\sigma\sigma -}$ given by \re{C-3d} start to deviate from their exact 
values  for $\epsilon>0.5$. We recall that the relations \re{C-3d} were obtained under the assumption that the levels $\Delta_\pm$ are separated
from the rest of the spectrum of $\Delta_i$ by a finite gap, e.g. $|\Delta_+-\Delta_-|\ll |\Delta_--\Delta_T|$. Since 
$\Delta_-(\epsilon)$ decreases linearly with $\epsilon$, this condition is not fulfilled  for $\epsilon\gg v$ and, as a consequence,
$C^2_{\sigma\sigma -}$ receives the additional corrections due to mixing of the operator $O_-$ with the stress-energy tensor. 
      
It is interesting to note that, according to \re{v},  the ratio of the OPE coefficients $C_{\sigma\sigma 2}/C_{\sigma\sigma 1}$ is 
anomalously small. This makes the properties of three-dimensional CFT's in the vicininity of critical 3d Ising model   similar to those 
of four-dimensional  $\mathcal N=4$ SYM theory in which case the OPE coefficients satisfy  \re{sol2} at large $N$. Notice that for the 
operators with sufficiently large scaling dimensions in unitary CFT's, their OPE coefficients have to fall off exponentially fast
\cite{Pappadopulo:2012jk}.  Yet another surprising feature of 3d Ising point is that the same asymptotic behaviour also holds for the 
lowest spin$-2$ operators.

We would like to emphasize that the above analysis relies on the assumption that the extremal
solutions living on the boundary of the allowed region of $\Delta_i$'s admit an interpretation in terms of 
three-dimensional CFT's, so that the curves shown in Fig.~\ref{fig3} can be interpreted as describing a continuous flow of the conformal data.~\footnote{I would like to thank Slava Rychkov for discussing
this point.}
This assumption is nontrivial and it does not hold e.g. for two-dimensional CFT's. In the latter case, the spectrum of the scaling dimensions of the extremal solutions
to the left of the two-dimensional Ising point is inconsistent with Virasoro symmetry  
making a CFT interpretation impossible~\cite{El-Showk:2014dwa}.
     
\subsection{Level-crossing in QCD}

As another example of level-crossing phenomenon, we consider the spectrum of scaling dimensions of composite operators that 
appeared in the study of the QCD evolution equations for three-particle distribution amplitudes~\cite{Braun:1999te,Belitsky:1999bf}. 
In the simplest case of baryon distribution amplitude of helicity $1/2$, these operators involve
covariant derivatives acting on three quark fields and  have
the following form
\begin{align}\label{bar}
B_{n_1n_2n_3}=\epsilon^{ijk} D^{n_1} q_i^{\uparrow} \, D^{n_2} q_j^{\downarrow}\,  D^{n_3} q_k^{\uparrow} \,,\qqqquad S=n_1+n_2+n_3\,,
\end{align}
where $q_i^{\uparrow(\downarrow)} =\frac12(1\pm \gamma_5) q_i$ are quark fields with definite chirality and with color $i=1,2,3$ corresponding
to the fundamental representation of the $SU(3)$ gauge group. 

As was shown in \cite{Braun:1999te},  the one-loop dilatation operator acting on the operators $B_{n_1n_2n_3}$ can be mapped into a Hamiltonian of a spin chain of length $3$  
\begin{align}\label{H-bar}
\mathbb H_{1/2} =  H_1 + \epsilon \,    H_2\,,
\end{align}
where spin operators at each site are identified with the generators of the collinear $SL(2)$ subgroup of the conformal group \footnote{  
Although the conformal symmetry of QCD is broken at quantum level, the one-loop dilatation operator respects the conformal
symmetry, see e.g. \cite{Braun:2003rp}. }
acting on
quark fields in \re{bar}.
Here $H_1$ coincides with a Hamiltonian of a completely integrable Heisenberg $SL(2)$ spin chain and the operator
$H_2$ is given by
\begin{align}
H_2=-\lr{ {1\over S_{12}^2} + {1\over S_{23}^2}}\,,
\end{align}
where $S_{ij}^2$ stands for the sum of two $SL(2)$ spins at sites $i$ and $j$. The eigenvalues of \re{H-bar} define the spectrum of anomalous
dimensions of baryonic operators \re{bar} at one loop.

The Hamiltonian \re{H-bar} depends on the parameter $\epsilon$. For baryonic operators \re{bar}, its value is uniquely fixed $\epsilon=1$.
To understand the properties of \re{H-bar}, it is convenient to treat $\epsilon$ as a new coupling constant and examine the dependence of 
the eigenvalues of \re{H-bar} on $\epsilon$. 
For $\epsilon=0$ the Hamiltonian \re{H-bar} is completely integrable and its exact eigenspectrum can be found with a help of Bethe
ansatz technique. For $\epsilon\neq 0$ the Hamiltonian \re{H-bar} is neither integrable, nor cyclic symmetric. Nevertheless, its
eigenspectum can be determined for abritrary $\epsilon$ using the technique described in \cite{Braun:1999te,Belitsky:1999bf}.

The Hamiltonian \re{H-bar} is invariant under the exchange of two sites $1\leftrightarrow 3$ and its eigenstates can be classified
according to the parity under this transformation. For $\epsilon=0$ the eigenstates with different parity have the same energy
in virtue of integrability, but the degeneracy is lifted for $\epsilon\neq 0$. The flow with $\epsilon$ of energy levels with different parity  is independent from one another and the non-crossing rule is not applicable. At the same time, the levels with the same parity are not
allowed to cross. 

\begin{figure}[t!bp]
\hspace*{-7mm} 
\includegraphics[width =.5\textwidth]{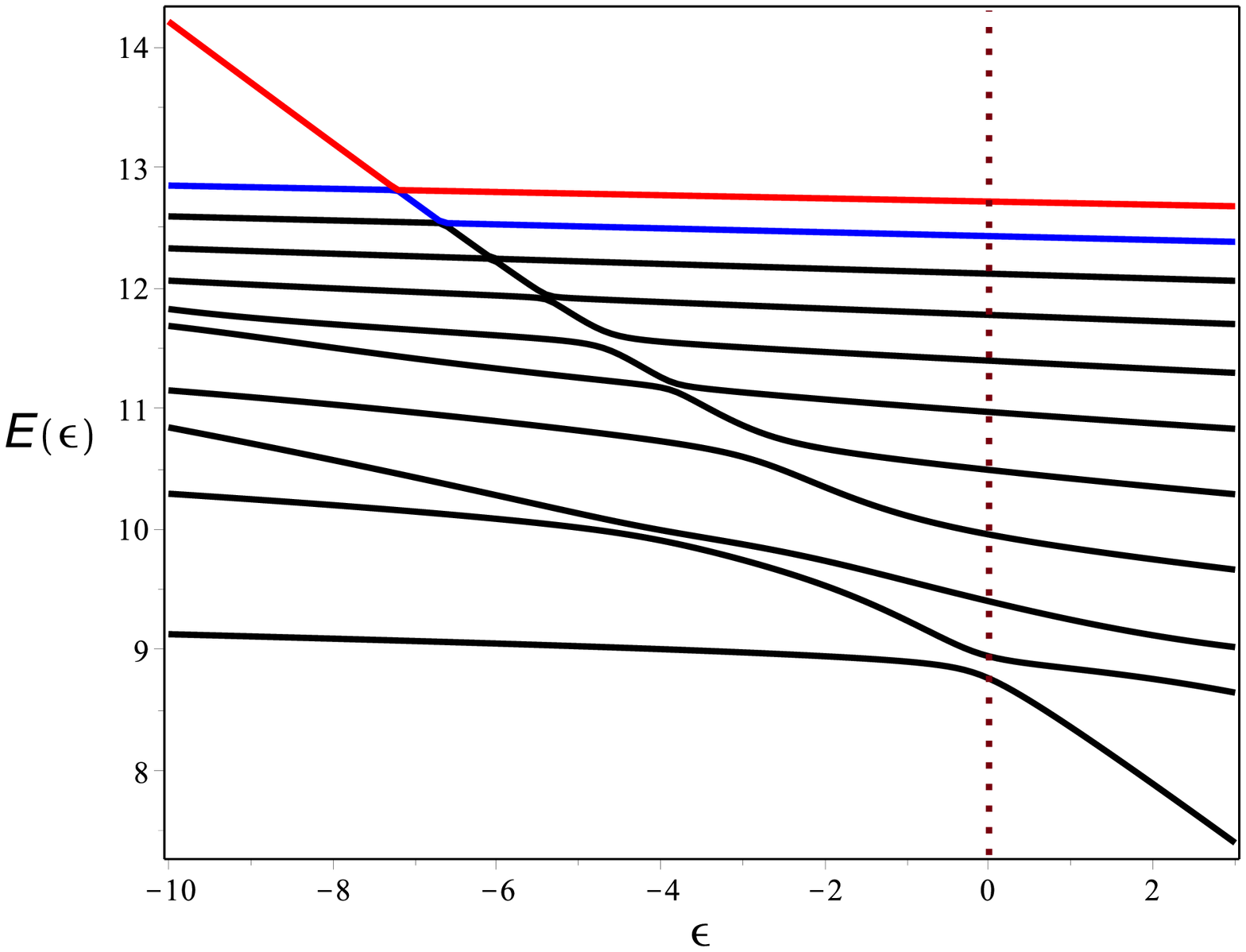} \qquad
 \includegraphics[width =.467\textwidth]{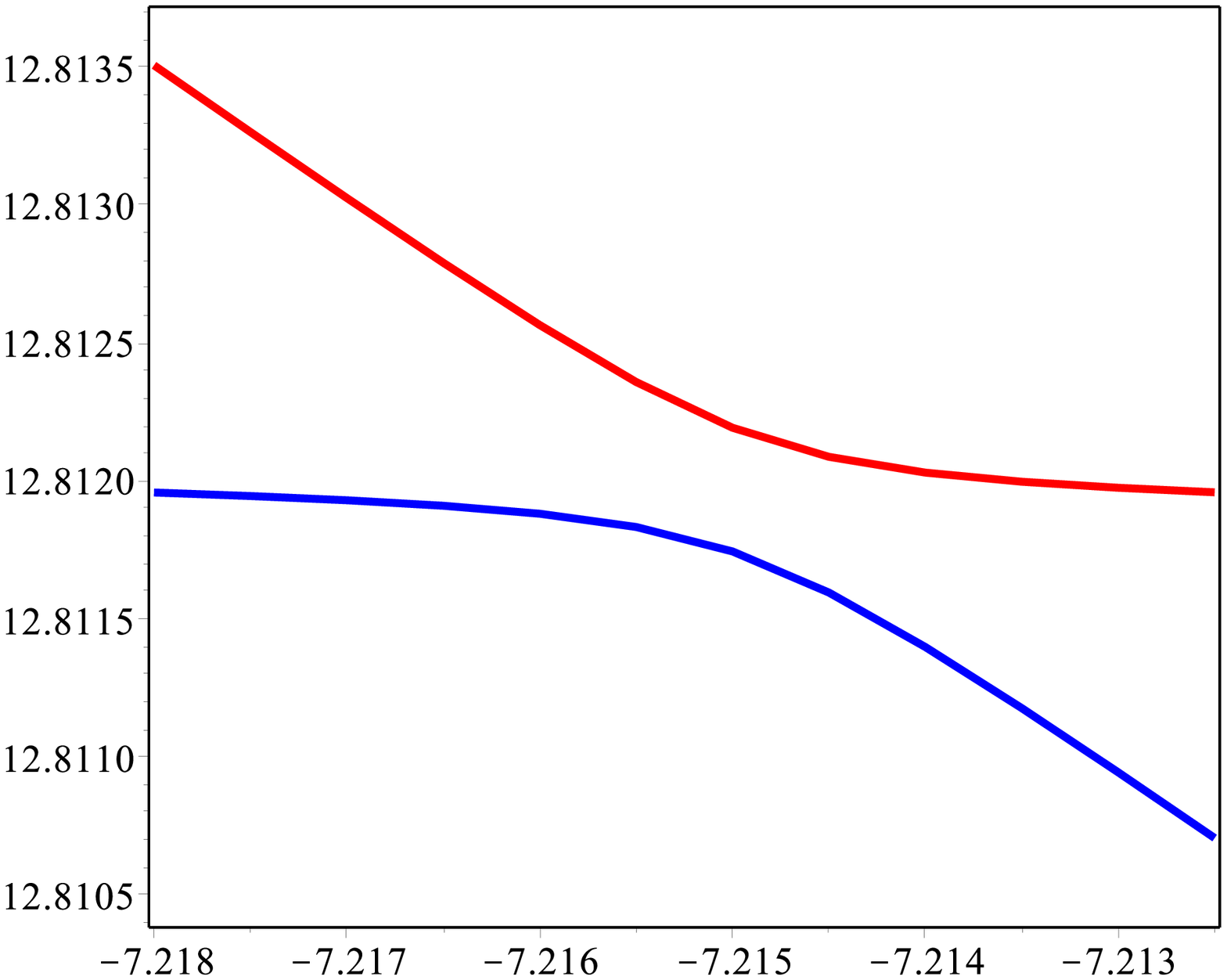} 
\caption{The flow of energies of parity-even eigenstates of the Hamiltonian \re{H-bar} for $S=20$. The vertical dotted line 
indicates the spectrum of $H_1$. The two highest levels are shown by red and blue lines. The
flow of energies close to the crossing point is zoomed in on the right panel.
}
\label{fig4}
\end{figure}   

As an example, we show on Fig.~\ref{fig4} the $\epsilon-$dependence of the energies of parity-even states
with the total spin $S=20$. We observe that for sufficiently large negative $\epsilon$ the energy levels approach each other.  
As was shown in \cite{Braun:1999te}, in the upper part of the spectrum, the crossing points correspond to the collision of  
levels of the Hamiltonians $H_1$ and $H_2$
\begin{align}\label{H1H2}
H_1\ket{\psi_1} =  e_1 \ket{\psi_1}\,,\qquad H_2\ket{\psi_2} = e_2\ket{\psi_2}\,,
\end{align}
with $\vev{\psi_1|\psi_2}\neq 0$ and $\vev{\psi_i|\psi_i}=1$. Assuming that the rest of eigenstates is irrelevant in the vicinity of
the crossing point,  we can look for the eigenstate of \re{H-bar} on a linear space spanned by the states $\ket{\psi_1}$ and $\ket{\psi_2}$.  Then, it is convenient to define the state 
\begin{align}
\ket{\hat\psi_1} = c\lr{\ket{\psi_1} - \ket{\psi_2} \vev{\psi_2|\psi_1}}\,,\qquad c = (1-|\vev{\psi_1|\psi_2}|^2)^{-1/2}\,,
\end{align}
such that $\vev{\hat\psi_1|\psi_2}=0$ and $\vev{\hat\psi_1|\hat\psi_1}=1$, and evaluate
the matrix elements of $\mathbb H_{1/2}$ with respect to $\ket{\hat\psi_1}$ and $\ket{\psi_2}$. In the standard manner, the diagonal matrix elements
define corrections to the energies of two states
\begin{align}\notag\label{E1E2}
 E_1(\epsilon) {}& = e_1+\epsilon\vev{\psi_1|H_2|\psi_1}+ O\left(\vev{\psi_1|\psi_2}^2\right)\,,
\\[2mm]
 E_2(\epsilon) {}& = \epsilon \,e_2 +  \vev{\psi_2|H_1|\psi_2} \,.
\end{align}
The off-diagonal matrix element defines their interaction energy
\begin{align}\label{v-int}
v= \vev{\hat\psi_1|\mathbb H_{1/2}|\psi_2} =c \vev{\psi_1|\psi_2} \left[ e_1 -\vev{\psi_2|H_1|\psi_2} \right]\,.
\end{align}
It is proportional to the overlap of the two eigenstates $\vev{\psi_1|\psi_2}$ and does not depend on $\epsilon$. As follows from \re{E1E2}, the levels could cross
at $ E_1(\epsilon)= E_2(\epsilon)$, or equivalently 
\begin{align}\label{eps-star}
  \epsilon_\star= { e_1  -  \vev{\psi_2|H_1|\psi_2} \over e_2- \vev{\psi_1|H_2|\psi_1} } \,. 
\end{align}
The crossing does not happen due to nonzero interaction energy $|v|\neq 0$. The eigenvalues of $\mathbb H_{1/2}$ in the vicinity 
of \re{eps-star} are given by
\begin{align}\label{pre}
E_\pm(\epsilon) = \frac12\lr{E_1(\epsilon) + E_2(\epsilon)} \pm \sqrt{ \frac14\lr{E_1(\epsilon) - E_2(\epsilon)}^2 + |v|^2   }\,.
\end{align}
To make this expression more explicit, we consider the limit of large spin $S\gg 1$. In this limit, $E_\pm(\epsilon)$ describe 
the one-loop anomalous dimensions of baryonic operators \re{bar} with large number of derivatives $S$.

The Schr\"odinger equations \re{H1H2} have been thoroughly studied in \cite{Braun:1999te}. For the eigenstates with large total
spin $S$, their energies   can be expanded in powers of $1/S$ leading to~\footnote{Expression for $e_1$ in this relation only holds 
in the upper part of the spectrum close to the maximal energy level, the general expression for $e_1$ can be found in \cite{Braun:1999te}.}
\begin{align}\notag\label{semi1}
{}& e_1 = 6 \ln S  - 3\ln 3 - 6\psi(2) + {6\over S} (2- \ell_1) + O(1/S^2) \,,
\\ 
{}& e_2 = -{1 \over (\ell_2+1)(\ell_2+2)}+ O(1/S^2)\,,
\end{align}
where $\psi(x)=(\ln\Gamma(x))'$ is Euler's digamma function and nonnegative integers $\ell_1$ and $\ell_2$ enumerate the energy levels.
The eigenstates with the maximal $|e_i|$ correspond to $\ell_1=\ell_2=0$. In the similar manner, for the matrix elements
of Hamiltonians with respect to $\ket{\psi_1(\ell_1)}$ and $\ket{\psi_2(\ell_2)}$ we have
\begin{align}\notag\label{vevs}
 \vev{\psi_1|H_2|\psi_1} = {}& O(1/S^2)\,,
\\[2mm]  
  \vev{\psi_2|H_1|\psi_2}  = {}&  4\ln S + C(\ell_2) + O(1/S)\,,
\end{align}
where $C(\ell_2)=- 6 \psi(2) + 6 \psi(\ell_2+2) -4 \psi(2\ell_2+4)+ {2/(\ell_2+2)} + {2/(2\ell_2+3)}$. Following \cite{Braun:1999te},
we can also compute the overlap of the eigenstates at large $S$
\begin{align}\label{semi3}
|\vev{\psi_1|\psi_2}|  =   \kappa \, S^{\,\ell_1/2+1/4} \e^{-S/2}  \,,
\end{align}
where $\kappa = \e^{\ell_2\sqrt{3}}  [{6\sqrt{\pi} \ell_1! (\ell_2+1)(\ell_2+2)/( 2^{\ell_1}(2\ell_2+3)})]^{-1/2} $.

Substituting the above relations into \re{v-int} and \re{eps-star} we obtain the following expressions for
the interaction energy
\begin{align}\label{v-pre}
| v | =2\kappa\, S^{\,\ell_1/2+1/4} \e^{-S/2}   \ln S\,,
\end{align}
and for the location of the crossing point  
\begin{align}\label{eps-pre}
{}& \epsilon_\star = - (\ell_2+1)(\ell_2+2)\left[2 \ln (S+1) +c_0(\ell_2)-{6\ell_1 +c_{1}(\ell_2)\over S}\right] + O(1/S^2)\,,
\end{align}
where $c_0 = - 3\ln 3  - 6 \psi(2) -C(\ell_2)$ and $c_1$ is independent on $\ell_1$.
We recall that the integers $\ell_1$ and $\ell_2$ enumerate  the colliding levels. 

We are now ready to compare the relations \re{pre}, \re{v-pre} and \re{eps-pre} with the exact energy spectrum of the Hamiltonian \re{H-bar}
shown in Fig.~\ref{fig4} for $S=20$. We remind that for $\epsilon=0$ we have $\mathbb H_{1/2} = H_1$ and  the energy levels 
are given by $e_1$, Eq.~\re{semi1}, for $\ell_1=0,1,\dots$ starting from the maximal one. For $\epsilon\neq 0$ away from the crossing
point, we find from \re{E1E2} and \re{vevs} that $E_1(\epsilon)=e_1 + O(1/S^2)$, so that the variation of the energy with $\epsilon$ is very
slow at large $S$. We observe from Fig.~\ref{fig4} that at large negative $\epsilon$ the maximal energy grows linearly with $(-\epsilon)$.
According to \re{E1E2}, \re{semi1} and \re{vevs}, it is given by 
\begin{align}
E_2(\epsilon) = -{\epsilon \over (\ell_2+1)(\ell_2+2)}+4\ln S + C(\ell_2) + O(1/S)\,,
\end{align}
for $\ell_2=0$. Going to smaller $(-\epsilon)$ we find that this level crosses subsequently the levels $E_1(\ell_1)$ for $\ell_1=0,1,2,\dots$.
According to  \re{eps-pre}, the position of the crossing point scales as $\epsilon_\star(\ell_1)=-4\ln S +12 \ell_1/S+\dots$, so that the
distance between two subsequent points scales as $\epsilon_\star(\ell_1+1)-\epsilon_\star(\ell_1)=12/S$. The interaction energy \re{v-pre}
defines the minimal distance between the colliding levels, $E_+(\epsilon_\star) -E_-(\epsilon_\star)  = 2|v|$. As follows from \re{v-pre}, this
distance is exponentially small at large spin $S$ and increases as $S^{\ell_1}$ with the level number. 

These results  are in a good 
agreement with properties of the exact spectrum shown in Fig.~\ref{fig4}. As an example, we consider the flow of  two highest levels shown 
by red and blue lines in Fig.~\ref{fig4}. The energies of these levels follow \re{pre}. They approach each other for $\epsilon_\star=-7.215$
and the level splitting is $2|v|=4.46\times 10^{-4}$. Applying \re{v-pre} and \re{eps-pre} for $\ell_1=\ell_2=0$ and $S=20$, we find
$\epsilon_\star=-7.229$ and $2|v|=4.32\times 10^{-4}$ which are close to the exact values. 

Notice that the flow of energy levels in the lower part of the spectrum in Fig.~\ref{fig4} is more complicated and it deviates from \re{pre}.
The reason for this is that the interaction between the lowest energy eigenstates is much stronger compared to that 
for the highest energy levels. Namely, the eigenstates in the lower part of the spectrum mix strongly with few other eigenstates and
finding their energies requires diagonalization of the mixing matrix whose size grows as $\ln S$ for large spin. 
This problem was solved in \cite{Braun:1999te,Belitsky:1999bf}.  
 
\section{Concluding remarks}

In this note, we examined the properties of conformal operators whose scaling dimensions approach each other 
for some values of the parameters of a unitary CFT  and satisfy von Neumann--Wigner non-crossing rule. We argued that the scaling dimensions of such 
operators and their OPE coefficients have a universal scaling behavior in the vicinity of the crossing point. 
We demonstrate that the obtained relations are in a good agreement with the known examples of the level-crossing phenomenon 
in maximally supersymmetric $\mathcal N=4$ Yang-Mills theory,   three-dimensional conformal field theories and QCD.

In gauge theories, the scaling dimensions do not verify the non-crossing rule in the planar limit, but 
the level crossing leads to breakdown of $1/N$ expansion. Namely, the nonplanar corrections develop singularities close to the crossing point
that have to be resummed to all orders in $1/N$. We argued that their resummation  
is very similar to resolving the mixing of operators in gauge theories at weak coupling with $1/N$ 
playing the role of the coupling constant. The resulting expressions 
for the scaling dimensions and the OPE coefficients, Eqs.~\re{fin} and \re{C-dif1}, respectively, 
remain finite in the transition region and obey the non-crossing rule. Away from this region their values coincide with those in the planar limit.

The scaling dimensions of twist-two operators are known to be increasing functions of the Lorentz spin $S$. For
sufficiently large values of the coupling constant and/or the spin, they cross the  scaling dimensions of the leading
twist-four double-trace operators. For small (large) spin $S$, the crossing happens at strong (weak) coupling. We used the
known results for Konishi operators $(S=0)$ and for twist-two operators with large spin ($S\gg 1$) to show that, due to 
the level crossing, the scaling dimensions and the OPE coefficients of twist-two and twist-four operators get swapped 
at the crossing point leading to a dramatic change of their asymptotic behaviour.~\footnote{Classification of the operators 
according to their twist becomes redundant at the crossing point, since the anomalous dimension of twist-two operator  approaches 
the value $\gamma_S=2$ and its twist increases to $4$.}
 
This effect is believed to play an important role in verifying the conjectured $S-$duality 
in $\mathcal N=4$ SYM~\cite{Beem:2013qxa,Beem:2013hha}. In particular, for finite $N$, the spectrum of the scaling dimensions should be invariant under the weak/strong coupling
duality transformation $h\to 1/h$ with $h=g^2/(4\pi)$. In the special case of Konishi operator, the $S-$duality implies the
existence of a nonperturbative operator $K'$ whose scaling dimension at weak coupling is related to that of the Konishi
operator as $\Delta_{K'}(h)=\Delta_K(1/h)$ \cite{Tseng:2002pe}. Let us examine the flow of the scaling dimensions $\Delta_{K'}(h)$ and $\Delta_K(h)$ 
and ignore, for the sake of simplicity, the mixing with other operators. Obviously, the functions $\Delta_{K'}(h)$ and $\Delta_K(h)$ 
could cross each other at $h=1$. As before, this can not happen due to a mixing between the operators $K$ and $K'$.
Replacing $\Delta_1=\Delta_K(h)$ and $\Delta_2=\Delta_K(1/h)$ in \re{fin}, we obtain the expressions for $\Delta_\pm(h)$ which
verify the non-crossing rule and are invariant under the $S-$duality transformation, $\Delta_\pm(h) = \Delta_\pm(1/h)$. Adding more operators 
does not change this result but makes the corresponding picture of
levels flow more complicated due to appearance of the additional crossing points.~\footnote{Similar to the Konishi operator, each operator $O_i$ is accompanied by a 
 nonperturbative $S-$dual operator $O_i'$. Then, resolving the mixing of the operators at the
 crossing points, we find that their  scaling dimensions are obtained by sewing together pieces of
 functions $\Delta_i(h)$ and $\Delta_i(1/h)$. The resulting functions are indeed invariant under the weak/strong
 coupling duality. } It would be interesting to test this mechanism using a
lattice formulation of $\mathcal N=4$ SYM \cite{Catterall:2014vka}.

\section*{Acknowledgements}
 
I would like to thank Fernando Alday for his collaboration at the early stage of this project and for his helpful
comments. I am grateful to Sheer El-Showk, Sergey Frolov, Miguel
Paulos, Slava Rychkov and Dima Volin for useful discussions. I am especially indebted to Miguel Paulos for sharing with me the data from 
Ref.~\cite{El-Showk:2014dwa}. This research was supported in part by the
French National Agency for Research (ANR) under contract StrongInt (BLANC-SIMI-4-2011). 
   
\bibliographystyle{JHEP} 

\providecommand{\href}[2]{#2}\begingroup\raggedright\endgroup

\end{document}